\documentstyle[11pt]{article}
\begin{document}
\begin{titlepage}
\begin{flushright}
SUSSEX-AST 95/1-1\\
astro-ph/9501086\\
(January 1995)\\
\end{flushright}
\begin{center}
\Large
{\bf Generalised Scalar Field Potentials and Inflation}\\
\vspace{.3in}
\normalsize
\large{Paul Parsons and John D. Barrow} \\
\baselineskip 24pt
\lineskip 10pt
\normalsize
\vspace{.6 cm}
{\em Astronomy Centre, \\ School of Mathematical and Physical Sciences,\\
University of Sussex, \\ Brighton BN1 9QH, U.~K.}\\
\vspace{.6 cm}
\end{center}
\begin{abstract}
\noindent
We investigate the range of inflationary universe models driven by scalar
fields possessing
a general interaction potential of the form $V(\phi) = V_0 \phi^n
\exp(-\lambda \phi^m)$. Power-law, de Sitter and
intermediate inflationary universes emerge as special cases together
with several new varieties of inflation. Analysing the behaviour of
these models at the extrema of $\phi$ we derive sufficient constraints
on the $m$ - $n$ parameter space such that inflation may occur as both an early
and late-time phenomenon. We also compute the scalar and tensor
perturbations produced in the model and compare these results with recent
observations.
\end{abstract}

\begin{center}
\vspace{1cm}
PACS number~~~98.80.Cq\\
\vspace*{4cm}
Submitted to {\bf Physical Review D}
\end{center}
\end{titlepage}

\baselineskip 24pt
\lineskip 10pt
\parskip=11pt

\def\theequation{\arabic{section}.\arabic{equation}}

\section{Introduction}

Inflationary cosmologies have become the most
popular models of the early universe. Their ability
to solve many of the problems inherent in the standard cosmology \cite{KT,LIN},
whilst simultaneously offering a mechanism for the generation of the seed
fluctuations for structure formation \cite{LL93}, consistent with recent
observations \cite{COBE,GOR}, make inflation a very elegant and
appealing scenario. Many variations now exist:
in addition to Guth's original version with an ephemeral cosmological
constant \cite{G81}, there exist models in which the
self-interaction potential $V$ for the {\it inflaton} field $\phi$
assumes exponential \cite{JJH,BB,SB90} or
positive power-law forms \cite{L83}. These
potentials give rise to power-law and polynomial-chaotic inflation
respectively \cite{L83,LM}.
Also of current interest, because of their ability to produce unusual
perturbation spectra, are the intermediate inflationary models of
[12 -- 15]. These arise when the potential behaves
asymptotically as a decaying power-law.

In this paper we endeavour to combine all of these models as special
cases of a single more general potential,
\begin{equation}
\label{pot1}
V(\phi) = V_0 \phi^n
\exp(-\lambda \phi^m)\,.
\end{equation}
This possibility was
originally examined by Barrow \cite{BE1}, where it was shown that $m
\leq 1$ was a {\em necessary} condition for inflation at large values
of $\phi$. In the absence of an exact solution for the potential
of Eq.~(\ref{pot1}) we
demonstrate that the slow-roll approximation \cite{LPB} is generically
satisfied at large $\phi$ in all such models, when $m \leq 1$. We employ
asymptotic techniques to ascertain the behaviour of
the system at large $\phi$ in all regions of the $m$ - $n$ parameter
space, revealing
some new types of inflationary universe. We are able to verify that $m
\leq 1$ is also a {\em sufficient} condition for the potential to
support an epoch of inflation at large $\phi$, and there are no constraints
on the value of $n$ in this limit.
We find that when $0 < m \leq 1$ inflation occurs as a late-time
phenomenon. However, in most cases with $m<0$ we find that $\phi$ grows
large as $t$
becomes {\em small} and inflation is an {\em
early} time feature. In these models $\dot{\phi}$ is negative and
the potential steepens
at small $\phi$ to end inflation naturally. Furthermore, if $m<0$, and
$V$ is an even function of $\phi$, then the potential will contain a minimum
at $\phi=0$, which allows the field ultimately to decay
into particles that reheat the universe once inflation ends \cite{KT},
without requiring
additional modifications to the model. We also find that when $m>1$,
inflation occurs uniquely at early times before the kinetic
energy of the field comes to dominate as $t \rightarrow \infty$.
If $0<m<1$ inflation proceeds indefinitely.
This also occurs in some models with $m<0$ and should be viewed as
describing the behaviour of the system on a discrete portion of some
grander potential with a minimum.
Finally, we calculate the scalar and tensor fluctuation spectra arising in
the generalised model. We use these results in conjunction with
observations to constrain further the allowed parameter values in
Eq.~(\ref{pot1}).

\section{Equations of Motion}
\label{eom}
\setcounter{equation}{0}

We shall concern ourselves with the behaviour of a zero-curvature Friedmann
universe. The matter content is dominated by a homogeneous
scalar field $\phi$ with potential
$V(\phi)$. The scale factor of the universe is $a(t)$, where $t$ is
synchronous cosmic time, and we define the Hubble expansion parameter as
\begin{equation}
H = \frac{\dot{a}}{a} \,,
\end{equation}
where an overdot represents a derivative with respect to $t$. Einstein's
equations (in units with $8\pi M_{Pl}^{-2} = c = 1$) are then
\begin{eqnarray}
\label{em1}
\ddot{\phi} + 3H\dot{\phi} & = & -V' \,, \\
\label{em2}
3H^2 & = & \frac{1}{2}{\dot{\phi}}^2 + V \,, \\
\label{em3}
\dot{H} & = & -\frac{1}{2}{\dot{\phi}}^2 \,,
\end{eqnarray}
where primes denote derivatives with respect to the scalar field and
$M_{Pl}$ is the Planck mass. Solution of these equations exactly is not
always possible, although we can often make progress within the context
of the slow-roll approximation. The Hubble Slow-Roll Approximation (HSRA)
\cite{LPB} is defined by the smallness of a set of parameters, the first
two of which are given by
\begin{eqnarray}
\epsilon_{\scriptscriptstyle H} & = & 2\left(\frac{H'}{H}\right)^2\:
 \left(\equiv \:3 \, \frac{\dot{\phi}^2/2}{V + \dot{\phi}^2/2}\right)
 \quad , \\
\eta_{\scriptscriptstyle H} & = & 2\frac{H''}{H}\: \left(\equiv \:- 3 \,
\frac{\ddot{\phi}}{3 H \dot{\phi}}\right)\,.
\end{eqnarray}
The condition $\epsilon_{\scriptscriptstyle H} <1$ is identical to the
condition
$\ddot{a}(t) > 0$ for inflation. In \cite{LPB} the HSRA was formulated as a
perturbative expansion, here we shall just be concerned
with the zeroth order approximation, which neglects the kinetic
terms (${\dot{\phi}}^2$ and $\ddot{\phi}$) in the equations of motion. In
this regime Eqs.~(\ref{em1}) - (\ref{em2}) simplify to
\begin{eqnarray}
\label{sr1}
3H^{2} & = & V\,, \\
\label{sr2}
3H\dot{\phi} & = & -V'\,.
\end{eqnarray}
Hence,
\begin{equation}
\label{sr3}
\dot{\phi} = -\frac{V'}{\sqrt{3V}}\,.
\end{equation}

For completeness we also mention the Potential Slow-Roll Approximation
(PSRA) \cite{LPB}, the validity of which is given by the smallness of the
parameters
\begin{eqnarray}
\epsilon_{\scriptscriptstyle V} & = & \frac{1}{2}\left(\frac{V'}{V}
                                      \right)^2\,,\\
\eta_{\scriptscriptstyle V} & = & \frac{V''}{V}\,.
\end{eqnarray}
The PSRA is in some respects more useful than the HSRA in that it allows
a certain amount of insight into the behaviour of the system simply
from examining the form of the potential. However, one
should exercise caution with such an approach.
Smallness of the PSRA parameters only classifies the flatness of the
potential; it carries no information regarding the initial conditions
$\{\phi_0, \dot{\phi}_0\}$ and so offers no guarantee that the field
will be genuinely slow-rolling. Thus for the PSRA to be valid requires
the additional assumption that the evolution has reached a form of
attractor solution, onto which all trajectories from all possible
initial conditions will have converged to within acceptable tolerances.
This is the inflationary attractor hypothesis discussed in \cite{LPB}.
The parameters $\epsilon_{\scriptscriptstyle V}$ and $\eta_{
\scriptscriptstyle V}$ for the potential of Eq.~(\ref{pot1}) are
\begin{eqnarray}
\label{epsv}
\epsilon_{\scriptscriptstyle V} & = & \frac{1}{2\phi^2}\left(n -
                                      \lambda m \phi^m\right)^2
                                      \,,\\
\label{etav}
\eta_{\scriptscriptstyle V} & = & \frac{1}{\phi^2}\left[n\left(
              \frac{n}{2}-1\right) -
              \lambda m\left(n+m-1\right)\phi^m + \frac{\lambda^2
              m^2}{2}\phi^{2m}\right]\,.
\end{eqnarray}
These expressions will prove useful in our later discussion of
perturbation spectra
as well as providing a reliable probe for determining when inflation
occurs.

\section{Asymptotic Analysis of ${\bf V(\phi) = V_0\phi^n \exp\left(-\lambda
\phi^m\right)}$}
\setcounter{equation}{0}

The behaviour of inflationary universe models driven by potentials of the form,
 $V(\phi) = V_0 \phi^{n}\exp({-\lambda \phi^{m}})$,
 was first investigated and classified by Barrow \cite{BE1}, where it
was shown that late-time inflation only occurs for models in which
$m \leq 1$, encompassing a broad spectrum of possibilities. When $m=0$
and $n$ is positive we have
polynomial-chaotic inflation, for negative $n$
the evolution asymptotes to intermediate inflation and when $m=1$,
$n=0$ we retrieve the power-law inflationary solution. The choice
$m=n=0$ leads back to Guth's original model \cite{G81} with exponential
expansion.

We shall now examine how this model behaves in the parameter ranges
$0<m<1$ and $m<0$. In these cases the equations of motion are not
exactly soluble, although we may make progress within the context of the
slow-roll approximation. Generically this will apply on the
asymptotic region of the potential, {\em ie} at large $\phi$, allowing
us to employ asymptotic techniques \cite{OLV} to solve Eqs.~(\ref{sr3})
and (\ref{pot1}) and obtain the complete slow-rolling solution.

\subsection{The case ${\bf 0<m<1}$}

Differentiating Eq.~(\ref{pot1}) with respect to $\phi$ yields
\begin{equation}
\label{vip}
V' = V_o\left(n-\lambda m \phi^{m}\right)\phi^{n-1}\exp(-\lambda \phi^{m})\,.
\end{equation}
We are working asymptotically in $\phi$, where the slow-roll conditions
are well satisfied and Eqs.~(\ref{sr1}) - (\ref{sr3}) are valid.
Substituting Eq.~(\ref{vip}) into Eq.~(\ref{sr3}) and taking the
asymptotic limit gives
\begin{equation}
\dot{\phi} = \sqrt{\frac{V_o}{3}} \lambda m
             \phi^{m + \frac{n}{2} -
             1}\exp{\left(-\frac{\lambda}{2}\phi^{m}\right)}\,.
\end{equation}
At large $\phi$ this expression integrates approximately to yield
\begin{equation}
\label{app1}
t(\phi) = \frac{2}{\lambda^2
m^2}\sqrt{\frac{3}{V_0}} \phi^{2-2m-\frac{n}{2}}
\exp\left(\frac{\lambda}{2}\phi^m\right)\,.
\end{equation}
In the absence of an exact inversion to $\phi(t)$ we express
Eq.~(\ref{app1}) as,
\begin{equation}
\left(2-2m-\frac{n}{2}\right)\ln{\phi} + \frac{\lambda}{2} \phi^{m} =
\ln \left[\sqrt{\frac{V_0}{3}} \frac{\lambda^2 m^2}
{2} t\right] \,.
\end{equation}
Asymptotically the power law in $\phi$ dominates the left hand side
and so the lowest-order approximation is
\begin{equation}
\label{pe1}
\phi(t) = \left(\frac{2}{\lambda}\right)^{\frac{1}{m}}\ln^{\frac{1}{m}}
          \left[\sqrt{\frac{V_0}{3}} \frac{\lambda^2
          m^2}{2}  t \right]\,.
\end{equation}
Rearranging Eq.~(\ref{app1}) and substituting Eq.~(\ref{pe1}) yields
$\phi(t)$ to second order,
\begin{equation}
\label{app2}
\phi(t) =
          \left(\frac{2}{\lambda}\right)^{\frac{1}{m}}\ln^{\frac{1}{m}}\left[
          \frac{\sqrt{\frac{V_0}{3}} \frac{\lambda^2 m^2}{2}
          t}{{\left(\frac{2}{\lambda}\right)}^{\frac{2}{m}-\frac{n}{2m}-2}
          \ln^{\frac{2}{m}-\frac{n}{2m}-2}\left[\sqrt{\frac{V_0}{3}}\frac{
          \lambda^2 m^2}{2} t\right]}\right] \,.
\end{equation}
We can now compute $H(t)$. From Eq.~(\ref{sr1}) we have
\begin{equation}
\label{happ}
H(\phi) = \sqrt{\frac{V_o}{3}}
          \phi^{\frac{n}{2}}\exp{\left(-\frac{\lambda}{2}
          \phi^{m}\right)}\,.
\end{equation}
Inserting Eq.~(\ref{app2}) into Eq.~(\ref{happ}) and taking the
large $t$ limit ($t \rightarrow \infty$ as $\phi \rightarrow
\infty$), we find
\begin{equation}
H(t)=\frac{1}{\lambda m^2}\left(\frac{2}{\lambda}\right)^
     {\frac{2-m}
     {m}}\frac{1}{t}\ln^{\frac{2-2m}{m}}\left[\sqrt{\frac{V_o}{3}}\frac{
     \lambda^2 m^2}{2} t\right]\,.
\end{equation}
This expression may be integrated to obtain the behaviour of the scale
factor. We thus obtain the full asymptotic time evolution of the system,
\begin{eqnarray}
\phi(t) & = & \left(\frac{2}{\lambda}\right)^{\frac{1}{m}}
              \ln^{\frac{1}{m}}t\,, \\
H(t) & = & \frac{1}{\lambda m^2}\left(\frac{2}{\lambda}\right)
           ^{\frac{2-m}{m}}\frac{1}{t}\ln^{\frac{2-2m}{m}}t\,, \\
a(t) & \propto & \exp\left[\frac{1}{\lambda m(2-m)}\left(\frac{2}
             {\lambda}\right)^{\frac{2-m}{m}}\ln^{\frac{2-m}{m}}t
             \right] \,.
\end{eqnarray}
Analysing the positivity of
$\ddot{a}(t)$ at large $t$ in the neighbourhood of $m=1$ confirms the
result that inflation only occurs when $m \leq 1$. Thus in general we
have a new form of inflation when $0<m<1,\: m \neq 1$.
As might be expected, the solution reduces to power-law inflation when
$m=1$ and is independent of $n$ to leading order.

\subsection{The case ${\bf m<0}$}
\label{nm}

The behaviour of $\dot{\phi}$ at large $\phi$ for $m<0$ leads to three
distinct cases:

\subsubsection{$m<0,\; n \neq 0,\,4$}
\label{gnm}

In this subclass of solutions the first term in Eq.~(\ref{vip}) will
dominate. Eqs.~(\ref{sr1}) and (\ref{sr2}) then imply
\begin{equation}
\label{phid1}
\dot{\phi} = -\sqrt{\frac{V_o}{3}}n \phi^{\frac{n}{2}-1}
             \exp \left(-
             \frac{\lambda}{2}\phi^{m}\right)\,.
\end{equation}
At large $\phi$ the integral of this expression is well approximated
by
\begin{equation}
\label{t1}
t(\phi) = \frac{1}{n}\sqrt{\frac{3}{V_0}}
          \left(\frac{2}{n-4}\right) \phi^{\frac{4-n}{2}}
          \exp\left(\frac{\lambda}{2}\phi^m \right)\,,
\end{equation}
which may be inverted asymptotically.
Substituting this result into Eq.~(\ref{happ}) allows
$H(t)$ to be approximated at large $\phi$, and integrated asymptotically
to obtain $a(t)$.
The full solution is
\begin{eqnarray}
\label{phi2}
\phi(t) & = & \left[\frac{n}{2}\sqrt{\frac{V_0}{3}}
              (n-4)t\right]^{\frac{2}{4-n}}\,,\\
\label{H2}
H(t) & = & \sqrt{\frac{V_0}{3}}\left[\frac{n}{2}
           \sqrt{\frac{V_0}{3}}(n-4)\right]
           ^{\frac{n}{4-n}}t^{\frac{n}{4-n}}\exp\left[-\frac{\lambda}
           {2}\left[\frac{n}{2}\sqrt{\frac{V_0}{3}}\right]
           ^{\frac{2m}{4-n}}t^{\frac{2m}{4-n}}\right]\,,\\
a(t) & \propto & \exp\left[\sqrt{\frac{V_0}{3}}\left(\frac{4-n}{4}\right)
                 \left[\frac{n}{2}(n-4)\sqrt{\frac{V_0}{3}}\right]
                 ^{\frac{n}{4-n}}t^{\frac{4}{4-n}}\gamma_1(t)\right]\,;
\end{eqnarray}
where $\gamma_1(t)$ is given by
\begin{equation}
\gamma_1(t) \equiv \exp\left\{-\frac{\lambda}{2}\left[\frac{n}{2}(n-4)\sqrt{
       \frac{V_0}{3}}\right]^{\frac{2m}{4-n}}
       t^{\frac{2m}{4-n}}\right\}\,,
\end{equation}
and we note that $\gamma_1(t) \rightarrow 1$ as $\phi \rightarrow \infty$.
Also,
$$H(t) \rightarrow \sqrt{\frac{V_0}{3}} \left[
     \frac{n}{2}(n-4)\sqrt{\frac{V_0}{3}}\right]
     t^{\frac{n}{4-n}}$$
for all $n$ as $\phi \rightarrow \infty$. Writing
\begin{equation}
\label{at}
a(t) \propto \exp\left[\alpha t^{\beta} \gamma_1(t)\right]\,,
\end{equation}
we find three subclasses, exhibiting the following asymptotic behaviour:

{\bf ({\it i}\,) ${\bf n>4}$}.  This implies $n(n-4) > 0$ and $\alpha,\:
\beta > 0$. Eq.~(\ref{t1}) reveals that $t \rightarrow 0$ as
$\phi \rightarrow \infty$ --- our asymptotic analysis is thus
concerned with the earliest stages of the universe's evolution.

{\bf ({\it ii}\,) ${\bf 0<n<4}$}. Since $n(n-4) < 0$, the constraints on
Eq.~(\ref{at}) in this case are $\alpha>0,\:1<\beta<\infty$.
Eq.~(\ref{t1}) implies $t \rightarrow -\infty$ as $\phi \rightarrow
\infty$ and inflation arises once again as a primordial effect
\footnote{
The apparent discrepancy in the definition of {\em early} between
cases {\bf ({\it i}\,)} and {\bf ({\it ii}\,)} ($t \rightarrow 0$ and
$t \rightarrow -\infty$) is a result of the special choices of
time co-ordinate arising in the particular solutions as a result of
ignoring integration constants. Both should be taken to represent
the early-time limit.}.

{\bf ({\it iii}\,) ${\bf n<0}$}. Here $n(n-4) >0$. The parameter constraints
are $\alpha>0,\:0<\beta<1$. In
this case we are concerned with the late-time evolution of the solution
since from Eq.~(\ref{t1}) $t \rightarrow
\infty$ as $\phi \rightarrow \infty$.
The solution asymptotes to an intermediate inflationary model.

Cases {\bf ({\it i}\,)} and {\bf ({\it ii}\,)} display inflationary
behaviour as an
early-time feature in a manner akin to chaotic inflation models with
concave potentials. Inflation proceeds for
a finite time before switching off as the potential becomes too steep.

\subsubsection{${\bf m<0,\; n=0}$}
\label{nmnn}

Eqs.~(\ref{sr1}) and (\ref{sr2}) imply
\begin{equation}
\dot{\phi} = \sqrt{\frac{V_0}{3}} \lambda m
             \phi^{m-1}\exp\left(-\frac{\lambda}{2} \phi^m\right)\,.
\end{equation}
In the asymptotic limit $t(\phi)$ becomes
\begin{equation}
\label{tn4}
t(\phi) = \sqrt{\frac{3}{V_0}} \frac{\phi^{2-m}}{
          \lambda m(2-m)}\exp\left(\frac{\lambda}{2}\phi^m\right)\,.
\end{equation}
Following the procedures of the previous section, we obtain the full
asymptotic solution,
\begin{eqnarray}
\label{rep1}
\phi(t) & = & \left[ \lambda m(2-m)\sqrt{\frac{V_0}{3}}
              \right]^{\frac{1}{2-m}}t^{\frac{1}{2-m}}\,,\\
H(t) & = & \sqrt{\frac{V_0}{3}}\exp\left[-\frac{\lambda}
           {2}\left\{\lambda m(2-m) \sqrt{\frac{V_0}{3}}
           \right\}^{\frac{m}{2-m}}t^{\frac{m}{2-m}}\right]\,,\\
\label{rep3}
a(t) & \propto & \exp\left[\sqrt{\frac{V_0}{3}} t\gamma_2(t)
                 \right]\,,
\end{eqnarray}
where,
\begin{equation}
\label{rep2}
\gamma_2(t) \equiv \exp\left[-\frac{\lambda}{2}\left\{\lambda m(2-m)\sqrt{
       \frac{V_0}{3}}\right\}^{\frac{m}{2-m}}t^{\frac{m}{2-m}}\right]\,.
\end{equation}
We see that $\gamma_2(t) \rightarrow 1$ as $\phi
\rightarrow \infty$ and so we recover the
de Sitter solution in this limit, as expected. We also see from
Eq.~(\ref{tn4}) that $t \rightarrow -\infty$ as $\phi
\rightarrow \infty$ and inflation is an early-time feature in the
evolution of the universe.

\subsubsection{${\bf m<0,\; n=4}$}

The behaviour of $\dot{\phi}$ at large $\phi$ for this model is
identical to the case for arbitrary $n$. We have
\begin{equation}
\dot{\phi} = -4\sqrt{\frac{V_0}{3}}\phi
             \exp\left(-\frac{\lambda}{2}\phi^m\right)\,.
\end{equation}
However, when this function is integrated in the asymptotic limit we
encounter qualitatively different behaviour, vested in the appearance of
a logarithmic factor,
\begin{equation}
t(\phi) = -\frac{1}{4}\sqrt{\frac{3}{V_0}}\left(\ln\phi
          \right)\exp\left(\frac{\lambda}{2}\phi^m\right)\,.
\end{equation}
Consequently, the asymptotic solution differs substantially,
\begin{eqnarray}
\label{phi3}
\phi(t) & = & \exp\left[-4\sqrt{\frac{V_0}{3}}t\right]\,,\\
H(t) & = & \sqrt{\frac{V_0}{3}} \exp\left[
           -8\left(\frac{V_0}{3}\right)^{\frac{1}{2}}t - \frac{\lambda}
           {2}\exp\left\{-4m\sqrt{\frac{V_0}{3}}t\right\}\right]\,,\\
\label{af}
a(t) & \propto & \exp\left\{-\frac{1}{8} \exp\left[-8
                 \sqrt{\frac{V_0}{3}} t\right] \gamma_3(t)\right\}\,,
\end{eqnarray}
where $\gamma_3(t)$ here takes the form
\begin{equation}
\gamma_3(t) \equiv \exp\left[-\frac{\lambda}{2}\exp\left\{-4m
       \left(\frac{V_0}{3}\right)^{\frac{1}{2}}t\right\}\right]\,,
\end{equation}
and tends to unity in the large $\phi$ limit. Eq.~(\ref{phi3}) reveals
that $t \rightarrow -\infty$ when $\phi \rightarrow \infty$; once again
we are dealing with the early evolution of the system and
Eq.~(\ref{af}) verifies the asymptotic approach of the model to standard
$V(\phi) \propto \phi^4$ models. Potentials such as this, which are
symmetric about $\phi=0$, contain minima and allow
a natural end to inflation within the model. One can thus expect a finite
amount of inflation in these theories, dependent on the initial value
of $\phi$.

\section{Analysis of ${\bf V(\phi)=V_0\phi^n\exp(-\lambda\phi^m)}$ at small
${\bf \phi}$}
\label{mb1}
\setcounter{equation}{0}

In many of the models examined in the previous section,
inflationary behaviour arose as an early-time feature at large $\phi$ values.
This poses the question of inflation at small $t$ in general for the
family of potentials defined by $V(\phi)=V_0\phi^n\exp(-\lambda\phi^m)$.
In \cite{BE1} a perturbative analysis was employed at large $t$ to
arrive at the constraint $m \leq 1$ necessary for inflation at late times
(corresponding to large $\phi$ for the particular
parameter ranges considered). A similar treatment here at small times
would be innapropriate because of the ambiguities concerning the
definition of ``early-time'' encountered in section \ref{nm}. Instead
we shall treat the scalar field as a time variable, as outlined in
\cite{L91}. The monotonicity of $\dot{\phi}$ that is required
to do this guarantees that each extremum of $t$ will correspond to a unique
extremum of $\phi$ and vice-versa; examining the system at
large and small $\phi$ is sufficient to obtain the complete solution
at early and late times. We thus complete our picture of inflationary models
arising from Eq.~(\ref{pot1}) by analysing their behaviour at small $\phi$.

Conditions on the potential such that inflation can occur, derived from
perturbation theory within this $\phi$-parameterised formalism concur
with those obtained from the PSRA
\footnote{The applicability of this
scheme at small $t$ could be regarded as questionable, since
the PSRA only applies generically once the system has had sufficient
time to settle into the inflationary attractor \cite{SB90,LPB}
. Here, we
assume that the slow-rolling portion of the potential is large enough to allow
the attractor to be attained well before leaving the early-time asymptopia,
although in practice this should be checked.}. We shall be concerned in
particular with $\epsilon_{\scriptscriptstyle V}$ since it offers a good
approximation to
$\epsilon_{\scriptscriptstyle H}$, the smallness of which is linked
directly to the
positivity of $\ddot{a}$. Inspecting
the form of $\epsilon_{\scriptscriptstyle V}$ for these models, given in
Eq.~(\ref{epsv}) we
see immediately that for small $\phi$, $\epsilon_{\scriptscriptstyle V}$ will
always blow
up unless $n=0$; $n=0$ is thus a necessary condition for inflation at
small $\phi$. In this case Eq.~(\ref{epsv}) simplifies considerably to
become
\begin{equation}
\epsilon_{\scriptscriptstyle V} = \frac{1}{2}\lambda^2m^2\phi^{2m-2}\,.
\end{equation}
Requiring this to be small as $\phi \rightarrow 0$ is equivalent to
demanding that $2m-2$ be positive or zero, leading to the constraint $m \geq
1$ for inflation. The bound in parameter space at small $\phi$ is thus
opposite to the bound at large $\phi$.

Next we look for small $\phi$ solutions.
Differentiating $V(\phi)$ and substituting into Eq.~(\ref{sr3}) yields
\begin{equation}
\label{phidsp}
\dot{\phi} = \sqrt{\frac{V_0}{3}}\lambda m\phi^{m-1}\exp\left(-
             \frac{\lambda}{2}\phi^m\right)\,.
\end{equation}
The choice $m=1$ leads to power-law inflation, and so we shall be
concerned only with $m>1$. Two distinct cases arise:

\subsection{The case ${\bf m>1,\:\neq2}$}

The analysis in this general case for positive $m$ at small $\phi$
parallels that of section \ref{nmnn} for negative $m$ at large $\phi$
and we obtain the approximate solution already given in
Eqs.~(\ref{rep1}) -- (\ref{rep2}).
When $1<m<2,\:t \rightarrow 0$ as $\phi \rightarrow 0$. If $m>2,\:
t \rightarrow -\infty$ as $\phi \rightarrow 0$ and the solution is valid at
early times. Also, $\gamma_2(t)$ tends to unity in
this limit, as expected.

\subsection{The case ${\bf m=2}$}

Here we discover a new type of behaviour. Eq.~(\ref{phidsp}) integrates
approximately at small $\phi$, giving
\begin{equation}
\label{tm2}
t(\phi) = \sqrt{\frac{3}{V_0}}\frac{1}{2\lambda}\left(\ln \phi\right)
          \exp\left(-\frac{\lambda}{2}\phi^2\right)\,.
\end{equation}
The full time-evolution, valid at small $\phi$, is then given by
\begin{eqnarray}
\phi(t) & = & \exp\left[2\sqrt{\frac{V_0}{3}}\lambda t\right]\,,\\
H(t) & = & \sqrt{\frac{V_0}{3}}\exp\left[-\frac{\lambda}{2}\exp\left(
           4\sqrt{\frac{V_0}{3}}\lambda t\right)\right]\,,\\
\label{lta}
a(t) & \propto & \exp\left[\sqrt{\frac{V_0}{3}}t\gamma_4(t)\right]\,,
\end{eqnarray}
with,
\begin{equation}
\gamma_4(t) \equiv \exp\left[-\frac{\lambda}{2}\exp\left(4\sqrt{\frac{V_0}{3}}
            \lambda t\right)\right]\,.
\end{equation}
Eq.~(\ref{tm2}) reveals that $t \rightarrow -\infty$ as $\phi \rightarrow
0$ and $\gamma_4(t) \rightarrow 1$.

Models with $m<1$ inflate at large $t$, whereas those with $m=1$ inflate
for all $t$ (power-law inflation). It should not be surprising then that
 the solutions inflate at small $t$ when $m>1$. Both the solutions
(\ref{rep2}) and (\ref{lta}),
presented above, tend to de Sitter expansion in the small $t$ limit.

\section{Density and Gravitational Wave Perturbations}
\setcounter{equation}{0}

A period of inflation in the early universe provides
a means to generate the small fluctuations from which
the large-scale structure we observe in the universe today can have grown. The
stretching and freezing-out of quantum excitations in the inflaton and
graviton fields during inflation gives rise to a spectrum of scalar
and tensor fluctuations in the cosmic microwave background; they can be
 classified
by the spectral indices $\hat{n}_s$ and $\hat{n}_g$ respectively \cite{LL93}.
These are defined as
\begin{eqnarray}
\hat{n}_s -1 & = & \frac{d \ln \delta^2_{{\scriptscriptstyle H}}(k)}{d
\ln k}\,,\\
\hat{n}_g & = & \frac{d \ln \sigma^2_{{\scriptscriptstyle H}}(k)}{d \ln k}\,,
\end{eqnarray}
where
\begin{equation}
\delta_{\scriptscriptstyle H}(k) = \frac{\delta \rho}{\rho}
\end{equation}
is the scalar density contrast on a scale corresponding to comoving
wavenumber $k=aH$ ($aH$ evaluated when the fluctuation crossed outside
the Hubble radius during inflation) and $\sigma_{\scriptscriptstyle
H}(k)$ is the
dimensionless strain
induced by the gravitational wave perturbations on scale $k$.
During slow-rolling inflation, one may express $\hat{n}_s$
and $\hat{n}_g$
to first order in the PSRA parameters as \cite{LL93}
\begin{eqnarray}
\label{ns}
1 - \hat{n}_s & = & 6\epsilon_{\scriptscriptstyle V} - 2\eta_
                    {\scriptscriptstyle V}\,, \\
\label{ng}
\hat{n}_g & = & -2\epsilon_{\scriptscriptstyle V}\,.
\end{eqnarray}
Furthermore, the ratio of the amplitudes of tensor to scalar modes in the
COBE signal on a scale corresponding to the $l^{th}$ multipole of the
temperature anisotropy expansion is
\begin{equation}
\label{R}
R_l =  \frac{25}{2}\epsilon_{\scriptscriptstyle V}\,,
\end{equation}
where $l$ is the multipole corresponding to the scale $k(\phi)$ at which
$\epsilon_{\scriptscriptstyle V}$ is evaluated.
Tensor fluctuations are thus always sub-dominant in slow-roll and
$\hat{n}_s$ is close to unity, as observed by COBE \cite{COBE}.

Much work has been carried out recently using the second-order
results of \cite{SL93}. However here the second-order
corrections to Eqs.~(\ref{ns}) -
(\ref{R}) prove too cumbersome to be of use, as well as exceeding
the accuracy of the approximate results already presented. Confining
ourselves to a first-order treatment and utilising Eqs.~(\ref{epsv}) and
(\ref{etav}) yields
\begin{eqnarray}
1-\hat{n}_s & = & \frac{2}{\phi^2}\left[n\left(\frac{n}{2}+1\right) + \lambda
            m\left(m-n-1\right)\phi^m + \frac{\lambda^2 m^2}{2}\phi^{2m}
            \right]\,,\\
\hat{n}_g & = & -\frac{1}{\phi^2}\left(n - \lambda m \phi^m\right)^2\,,\\
R_l & = & \frac{25}{4\phi^2}\left(n-\lambda m\phi^m\right)^2\,,
\end{eqnarray}
These expressions are evaluated at horizon crossing when $k(\phi) = aH$. The
relation between the field $\phi$ and scale $k$ is established by
considering the number of e-foldings of contraction
experienced by the comoving Hubble length $(aH)^{-1}$, given by
\begin{equation}
\bar{N}(k) \equiv \ln \frac{(aH)_f}{(aH)_i}\,,
\end{equation}
where the subscripts $i$ and $f$ denote the initial and final values of
$(aH)$.
The functional form of $\bar{N}(k)$ is
irrelevant for our discussion here although an expression is given in
\cite{LPB}. More pertinent here is the counterpart function,
$\bar{N}(\phi)$, given to lowest order in PSRA parameters by \cite{LPB}
\begin{equation}
\bar{N}(\phi_1,\,\phi_2) = -\sqrt{\frac{4\pi}{M^2_{Pl}}}\int^{\phi_2}_
                           {\phi_1}\frac{1}{\sqrt{\epsilon_
                           {\scriptscriptstyle V}(\phi)}}\,
                           d\phi\,.
\end{equation}
With a view to numerical evauation, the $M_{Pl}$ factors have been
restored to this expression and are retained hereafter.
The form of $\bar{N}$ is dependent upon the range of $m$ and $n$ and the
appropriate limit of $\phi$ required to obtain inflationary behaviour:

{\bf ({\it i}\,)} $0<m<1$ and $m\leq0,\,n=0$ as $\phi
\rightarrow \infty$ or $m>1$ as $\phi \rightarrow 0$. In
these cases we obtain,
\begin{equation}
\label{N1}
\bar{N}(\phi_1,\,\phi_2) \rightarrow \frac{1}{\lambda m(2-m)}
                         \frac{8\pi}{M^2_{Pl}}
                         \left[\phi^{2-m}_2 -\phi^{2-m}_1\right]\,.
\end{equation}

{\bf ({\it ii}\,)} $m \leq 0,\,n\neq0,\:\phi \rightarrow \infty$. The result
here is simpler,
\begin{equation}
\label{N2}
\bar{N}(\phi_1,\,\phi_2) \rightarrow \frac{1}{n}
                         \frac{8\pi}{M^2_{Pl}}
                         \left[\phi^2_1 - \phi^2_2\right]\,.
\end{equation}
Eqs.~(\ref{N1}) and (\ref{N2}), together with $\bar{N}(k)$ establish
the $\phi$ - $k$ correspondences
necessary to relate quantum fluctuations at a ``time'' $\phi$ to microwave
background fluctuations on a scale $k$.

A successful model of inflation requires a minimum of $70$ e-foldings
of comoving contraction (to solve the horizon and flatness problems)
before the conditions $\epsilon_{\scriptscriptstyle V}$,
$\eta_{\scriptscriptstyle V} \ll 1$ become violated
\cite{G81}. This constraint provides bounds on the potential in models
where $\phi_1$ is fixed by particular initial conditions.
When the potential possesses a minimum this does not
apply; then, $\epsilon_{\scriptscriptstyle V},\,\eta_{\scriptscriptstyle V}
\rightarrow 0$ at late times, and such
models must be modified if inflation is to end. Potentially, this
permits an enormous amount of expansion.

It was seen earlier that the amplitude of the fluctuations arising
from slow-rolling inflation is dominated by the scalar component. This
has been calculated to zeroth order in the PSRA \cite{LL93} as
\begin{equation}
\delta^2_{\scriptscriptstyle H} = \frac{512\pi}{75M^6_{Pl}}\frac{V^3}{V'^2}\,.
\end{equation}
For the potential of Eq.~(\ref{pot1}) this
implies
\begin{equation}
\label{delt}
\delta_{\scriptscriptstyle H} \simeq \frac{16\sqrt{2}}{5M^3_{Pl}}
                 \sqrt{\frac{V_0}{3}}\frac{
                 \phi^{\frac{n}{2}+1}\exp\left(-\frac{\lambda}{2}\phi^m
                 \right)}{\left|n-\lambda m\phi^m\right|}\,.
\end{equation}
The form of $\bar{N}(k)$ given in \cite{LPB} implies that $(aH)^{-1}$
shrunk to scales of astrophysical interest at approximately
$\phi=\phi_{\scriptscriptstyle{*}}$, 60 comoving e-folds before the
end of inflation, a figure which is
weakly sensitive to the physics of reheating. Models in which inflation
ends naturally (namely those occupying the regions
$m<0,\,n\geq0$ or $m>1,\,n=0$ of $m$ - $n$ parameter space) can thus
be constrained further by demanding $\delta_{\scriptscriptstyle H}\left(k_
{\scriptscriptstyle{*}}\right)$ be consistent with observational data.
This requires
knowledge of $\phi(\epsilon_{\scriptscriptstyle V})$ so that one may
obtain a reliable
approximation to $\phi_{end}\simeq\phi(\epsilon_{\scriptscriptstyle V}=1)$,
the
value of $\phi$ at which inflation ends. In particular, if $n=0$ we find
\begin{equation}
\phi_{end} = \left(\frac{16\pi}{\lambda^2 m^2M^2_{Pl}}\right)^{\frac{1}
             {2m-2}}\,.
\end{equation}
Which, in conjunction with Eq.~(\ref{N1}) yields
\begin{equation}
\phi_{\scriptscriptstyle{*}}^{2-m} = \left(\frac{16\pi}{\lambda^2
                                      m^2M^2_{Pl}}\right)^{\frac{2-m}{2m-2}}
                                      + \frac{15\lambda mM^2_{Pl}}{2\pi}
                                      (m-2)\,.
\end{equation}
Substituting this result, and $n=0$, into Eq.~(\ref{delt})
we obtain the model prediction,
\begin{equation}
\left[\frac{5\lambda m M^3_{Pl}}{16\sqrt{2}}\sqrt{\frac{3}{V_0}}
\delta_{\scriptscriptstyle H}\right]^{\frac{2-m}{1-m}} \approx
                            \left(\frac{16\pi}{
                            \lambda^2 m^2M^2_{Pl}}\right)^{\frac{2-m}{2m-2}}
                            + \frac{15\lambda mM^2_{Pl}}{2\pi}(m-2)\,.
\end{equation}
Under particular circumstances this expression simplifies. When $m=2$,
\begin{equation}
\delta_{\scriptscriptstyle H} \approx
                              \frac{4}{5M^2_{Pl}}\sqrt{\frac{2V_0}
                              {3\pi}}\,,
\end{equation}
and as $m \rightarrow \pm \infty$,
\begin{equation}
\delta_{\scriptscriptstyle H} \rightarrow \frac{16}{5\lambda m M^3_{Pl}}
                              \sqrt{\frac{2V_0}{3}}\left[\frac{15}{2\pi}
                              \lambda m^2 M^2_{Pl}\right]^{\frac{1-m}
                              {2-m}}\,.
\end{equation}
The most recent analysis of the COBE data \cite{GOR} gives
$\delta_{\scriptscriptstyle H}
\simeq 2.3\times 10^{-5}$ which, once $\lambda$ and $m$ have been fixed,
imposes a constraint on the value of $V_0$.

\section{Conclusions}
\setcounter{equation}{0}

We have studied the behaviour of inflationary universe models emerging
when one posits a form for the inflaton potential containing both
exponential and power-law factors. By examining the validity of the
slow-roll approximation in these models at
the extremes of $\phi$ we have determined the structure of the $m$ -
$n$ parameter space at early and late times. We have isolated the regions in
which inflation may occur in these two eras and we have obtained a wide
variety of qualitatively different
solutions to the equations of motion, dependent to a large extent on the
value of $m$. We find that when $m>1$, quasi-de Sitter inflation is
manifest at early-times
provided $n=0$ but this is not a feature as $t \rightarrow \infty$,
where the potential tends to zero faster than the kinetic energy of the
field, and all models are non-inflationary. This is
the case at all times when $m>1,\:n\neq 0$; these models can never
inflate. Within the broad range $0\leq m\leq 1$ inflation is generic
at late times for all $n$. In particular, when $m=0$, the solutions exhibit
polynomial-chaotic or intermediate inflationary behaviour (dependent
upon the sign of $n$) and, if $n=0$, we recover the de Sitter
solution. If $m=1$ and $n=0$ we have power-law inflation; when
$n\neq 0$ the behaviour will asymptote to the power-law form at large
$t$. When $0<m<1$ the scale factor approaches a new type of behaviour,
proportional to $\exp(\ln^{(2-m)/m}t)$ as $t \rightarrow \infty$. If
$m<0$ and $n\geq 0$, inflation occurs at early times; but, if $n<0$, it
proceeds once more as a late-time phenomenon and the precise form of the
inflationary behaviour is determined by the specific value of $n$.

In summary, our analysis extends the treatment presented in \cite{BE1} to
establish a set of sufficient conditions for inflation to occur at early
or late epochs, summarised in Table 1 below.


\begin{center}
\begin{tabular}{||c|c|c||} \hline
Early & Late & All \\
Times & Times & Times \\ \hline \hline
$m\leq 0,\:n\geq 0$ & $0\leq m\leq 1$ & $m=1,\: n=0$ \\
{\bf or} & {\bf or} &  {\bf or} \\
$m\geq 1,\:n=0$ & $m\leq 0,\:n<0$ & $m=n=0$ \\
 & & \\ \hline
\end{tabular}
\end{center}
\begin{center}


Table 1: The conditions on $m$ and $n$ in $V(\phi)=V_0\phi^n
\exp(-\lambda \phi^m)$\\ for inflation to occur.
\end{center}

The calculation of the perturbation spectra produced in these theories
allows further constraints to be placed on $m$, $n$ and $\lambda$ by
comparison with microwave background observation. Moreover, the
restrictions imposed on
$V_0$ from the amplitude of scalar fluctuations go some way toward
determining the energy scale at which inflation occurs \cite{LES}.

\section*{Acknowledgements}

PP is supported by a PPARC Postgraduate Studentship and JDB by a PPARC
Senior Fellowship.
\frenchspacing

\end{document}